# Stable Oxide Nanoparticle Clusters Obtained by Complexation

**J.-F. Berret**[@]
Matière et Systèmes Complexes, UMR 7057 CNRS Université Denis Diderot Paris-VII, 140 rue de Lourmel, F-75015 Paris France

**A. Sehgal, M. Morvan,**
Complex Fluids Laboratory, CNRS - Cranbury Research Center Rhodia Inc.,
259 Prospect Plains Road, Cranbury NJ 08512 USA

**O. Sandre,**
Laboratoire Liquides Ioniques et Interfaces Chargées, UMR 7612 CNRS Université Pierre et Marie Curie Paris-VI, 4 place Jussieu, Boîte 51, F-75252 Paris Cedex 05 France

**A. Vacher, M. Airiau**
Rhodia, Centre de Recherches d'Aubervilliers, 52 rue de la Haie Coq, F-93308 Aubervilliers Cedex France

**Abstract :** We report on the electrostatic complexation between polyelectrolyte-neutral copolymers and oppositely charged 6 nm-crystalline nanoparticles. For two different dispersions of oxide nanoparticles, the electrostatic complexation gives rise to the formation of stable nanoparticle clusters in the range 20 – 100 nm. It is found that inside the clusters, the particles are "pasted" together by the polyelectrolyte blocks adsorbed on their surface. Cryo-transmission electronic microscopy allows to visualize the clusters and to determine the probability distributions functions in size and in aggregation number. The comparison between light scattering and cryo-microscopy results suggests the existence of a polymer brush around the clusters.

## I - Introduction

Since the pioneering works by Kataoka and Harada [1,2], it has been recognized that attractive interactions between polyelectrolyte-neutral diblock copolymers and oppositely charges species result in the formation of new type of colloids. These colloids called polyion complex micelles [3] or colloidal complexes [4] form spontaneously by electrostatic self-assembly with a core-corona microstructure. With polyelectrolyte-neutral copolymers, the complexation is controlled by the appropriate choice of the polymer, its molecular weight and by the molecular weight ratio between the two blocks. Optimal conditions for complexation have been determined experimentally. These conditions are reached when the degree of polymerization of the neutral block is 2 - 5 times that of the charged block. So far, the specimens examined with respect to copolymer complexation comprise synthetic [5] and biological [1,6] macromolecules, multivalent counterions [7,8], surfactant micelles [4,9-11]. The formation of the mixed aggregates is generally understood as the result of a nucleation and growth mechanism of a microphase made from the oppositely charged constituents. This growth is arrested at a size which is fixed by the dimension of the polymer. Other complexation approaches using nanoparticles and polymers have also received attention recently [12-14].

In this communication, we report that the electrostatic complexation is effective with crystalline nanoparticles of size less than 10 nm. In order to demonstrate the generality of the approach, two different types of nanoparticles were considered. We have investigated anionically modified dispersions of cerium and iron oxide nanoparticles stabilized by citric acid. Using cryo-transmission electronic microscopy (cryo-TEM), we are taking advantage of the strong electronic contrast of the metallic atoms to visualize directly the mixed polymer-nanoparticle aggregates. We show that the particles are aggregated in densely packed clusters of size 20 – 100 nm, and we demonstrate that the clusters are surrounded by a neutral polymer corona. We also derive the probability distribution functions of aggregation numbers (*i.e.* the number of nanoparticles per cluster) and discuss the mechanisms of cluster formation.

## II - Experimental

The first system investigated in this work is a dispersion of cerium oxide nanocrystals (nanoceria) provided to us by Rhodia. The cerium oxide nanoparticles ($\rho$ = 7.1 g·cm$^{-3}$) suspensions were synthesized as cationic fluorite-like nanocrystals in nitric acid at pH 1.4. The synthetic procedure involves thermohydrolysis of an acidic solution of cerium-IV nitrate salt at high temperature, resulting in homogeneous precipitation of a cerium oxide nanoparticle pulp [15]. The size of the particles was controlled by addition of hydroxide ions during the thermohydrolysis. High resolution transmission electron microscopy have shown that the nanoceria consist of isotropic agglomerates of 2 - 5 crystallites with typical size 2 nm and faceted morphologies, and wide-angle x-ray scattering confirmed the crystalline fluorite structure of the nanocrystallites [15]. Cryo-TEM images of single nanoparticles are displayed in Fig. 1 (top three photographs). An image analysis performed on 350 particles allowed to determine the probability distribution function (pdf) in size for the nanoceria. This distribution was found to be well-accounted for by a log-normal function, with a most probable diameter $D_0[\gamma\text{-CeO}_2]$ = 6.9 ± 0.3 nm and a the polydispersity s = 0.15 ± 0.03. From this distribution, the radius of gyration and the hydrodynamic diameter were calculated [16] to be $R_G[\text{CeO}_2]$ = 3.5 nm and $D_H[\text{CeO}_2]$ = 8.6 nm, in good agreement with the direct x-ray and light scattering determinations (which are at 3.5 nm and 9.8 nm, respectively [15]).

The second dispersion investigated contains superparamagnetic nanoparticles of maghemite ($\gamma$-Fe$_2$O$_3$). The iron oxide nanoparticles ($\rho$ = 5.1 g·cm$^{-3}$) were synthesized by alkaline co-precipitation of iron II and iron III salts [17] and sorted according to size by successive phase separations [18]. The size distribution was monitored by vibrating sample magnetometry [19] and cryo-TEM experiments (image analysis on 470 particles). In both cases, the size distribution could be represented by a log-normal function with a most probable diameter $D_0[\gamma\text{-Fe}_2O_3] = 6.3 \pm 0.3$ nm and a polydispersity of $0.23 \pm 0.03$. Instances of single maghemite nanoparticles are illustrated in Fig. 1 (bottom). As for nanoceria, $R_G$ and $D_H$ were computed from this distribution and they were found in good agreement with the data obtained by neutron and light scattering (here, $R_G[\gamma-\text{Fe}_2O_3] = 3.1$ nm and $D_H[\gamma-\text{Fe}_2O_3] = 11$ nm [20]).

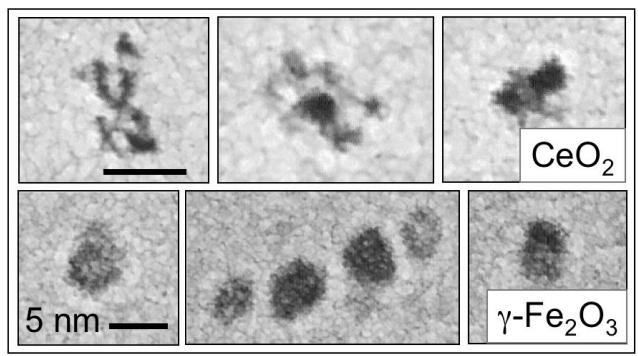

*Figure 1 :* Cerium (top) and iron (bottom) oxide nanoparticles as observed by cryogenic transmission electron microscopy. The stability of the dispersions is ensured by electrostatic interactions mediated by citrate ligands adsorbed on the surface. Most probable diameters are $D_0 = 6.9$ and $6.3$ nm, respectively.

At pH 7 - 8 the particles are stabilized by electrostatic interactions which are mediated by ligands (citric acid) adsorbed on the surface of the particles. For the two systems, we verified using ζ-potential measurements that citrate-coated nanoparticles are negatively charged (ζ = - 40 mV for nanoceria and ζ = - 32 mV for maghemite) and opposite in sign with respect to the polyelectrolyte block used for complexation. The light scattering and zeta potential characterisations of particles and polymer-particle aggregates were performed using the Zetasizer Nano ZS from Malvern Instruments.

The two types of nanoparticles described above were complexed with a polyelectrolyte-neutral diblock referred to as poly(trimethylammonium ethylacrylate)-*b*-poly(acrylamide). These copolymers, abbreviated as PTEA(5K)-*b*-PAM(30K) in the following were synthesized by controlled radical polymerization according to MADIX technology [21]. The chemical formulae of the monomers are given in Refs. [10,20]. PTEA is a strong polyelectrolyte and as such its 19 monomers (average) are fully ionized. The monomers are positively charged at all pH-values. The same diblock copolymers have been utilized for complexation with anionic surfactant micelles [11,22]. Polymer-nanoparticle complexes were obtained by simple mixing of stock solutions prepared at the same weight concentration (c = 0.1 – 1 wt.%) and pH (pH 7 – 8). The relative amount of each component is monitored by the mixing ratio X, which is defined as the ratios of the volumes of nanoparticle solution added relative to the polymer solution. The solutions were prepared at the preferred mixing ratio $X_P$ (here, $X_P = 1$ for both systems) *i.e.* at the ratio where all the components present in solution associate to form complexes. Protocols for mixing oppositely charged species in solutions have been described previously [22,23].

## III – Results and Discussion

Cryo-TEM was performed on mixed polymer-nanoparticle solutions for $CeO_2$ and $\gamma$-$Fe_2O_3$ and the results illustrated in Figs. 2 to 3 respectively. The photographs cover spatial fields that are approximately 0.6×0.4 $\mu m^2$ and display clusters of nanoparticles. Large visual fields are here shown in order to emphasize that the clusters are well-dispersed, a result which is consistent with the direct observations of the solutions (no separation). For these two samples, dynamic light scattering (DLS) reveals a slightly polydisperse diffusive relaxation mode in the scattered light autocorrelation function, associated with hydrodynamic diameters $D_H^{DLS}[CeO_2]$ = 75 nm and $D_H^{DLS}[\gamma - Fe_2O_3]$ = 70 nm. The values of the $\zeta$-potential for the two aggregates are around zero (± 5 mV), indicating that the positive and negative charges have been compensated in the complexation. Note that in the cluster state, the nanoparticles are stable over a broad range of pH (pH 3 – pH 10) and of ionic strengths (up to 0.2 M NaCl). The details of the colloidal stability of these hybrid colloids will be examined in a forthcoming paper.

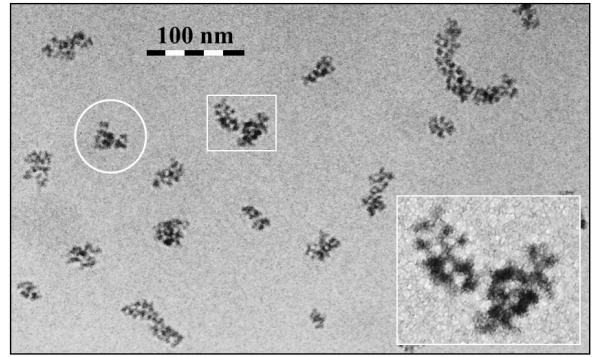

*Figure 2 : Cryo-TEM images of mixed nanoparticle/polymer complexes obtained using PTEA-b-PAM diblock copolymers and $D_0$ = 6.9 nm cerium oxide nanoparticles. The total concentration is c = 0.2 wt. % and X = $X_P$ (see text). Inset : zoom of two aggregates revealing the internal structure of the clusters. The white circle indicates the hydrodynamic size as determined by dynamic light scattering.*

In the cryo-TEM images, magnification of selected clusters (shown in squared insets) allow us to better distinguish the nanoparticles inside the aggregates. A closer inspection of Fig. 2 and 3 reveals that the aggregates are slightly anisotropic, with their largest dimensions comprised between 20 and 100 nm. For the analysis of the cluster morphology, it is assumed that the clusters can be represented by ellipsoids of revolution, and that their bidimensional projections are ellipses with a major and a minor axis, noted *a* and *b* respectively. Based on the image analysis of nearly 200 aggregates, the pdf's for the minor and major axis were obtained and the equivalent hydrodynamic diameter (maximum of the intensity distribution in DLS [16]) determined for each cluster population [20]. We found $D_H^{TEM}[CeO_2]$ = 45 ± 5 nm and $D_H^{TEM}[\gamma - Fe_2O_3]$ = 40 ± 5 nm. These values are lower by ~ 30 nm compared to the actual hydrodynamic diameter measured by light scattering for the same solutions. The difference between the light scattering and electron microscopy data may suggest the existence of a polymer brush around the inorganic clusters. Note that the brush is not detected using microscopy because of the weak electronic contrast of the polymers with respect to water. Typical sizes for the whole polymer-nanoparticle aggregates are shown by circles in Figs. 2 and 3.

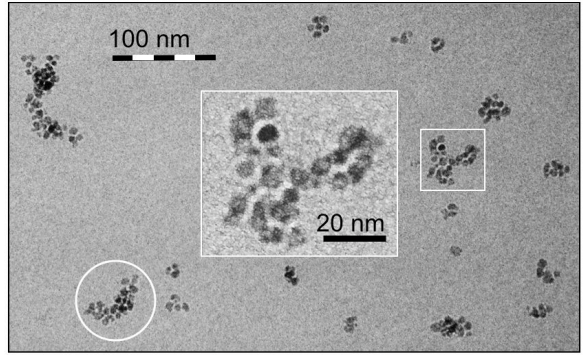

***Figure 3*** : *Same as in Fig. 2, but for clusters obtained using $D_0 = 6.3$ nm iron oxide nanoparticles.*

From the cluster size and distributions, we then derive the corresponding pdf's of aggregation numbers, as well as the number and weight average aggregation numbers $\overline{N}_n$ and $\overline{N}_w$ related to these distributions. The estimations of the aggregation numbers have been carried out by assuming that the clusters have an effective particle volume fraction $\phi = 0.3$ [20]. The aggregation numbers are then computed according to the expression $N = \phi \, ab^2 / \overline{D^3}$, where the upper bar refers to the average over the single particle distribution. This model is qualitatively consistent with results estimated by counting the particles directly on the cryo-TEM images. In Fig. 4, the aggregation number pdf's obtained for cerium and iron nanoparticles are plotted in a semi-logarithmic representation as a function of the reduced quantity $N/\overline{N}_n$. Both exhibit rather broad distributions which are approximated by an exponential function of the form :

$$P_n(N) \sim \exp(-N/\overline{N}_n) \qquad (1)$$

In the statistical arrays considered here, single unassociated particles were not observed and so Eq. 1 holds only for $N > 1$. From best fit calculations in Fig. 4, we found $\overline{N}_n = 25 \pm 4$ ($CeO_2$) and $16 \pm 3$ ($\gamma$-$Fe_2O_3$), with polydispersities $\overline{N}_w/\overline{N}_n = 1.7$ and $2.1$ respectively. At first, the existence of an exponential pdf for the aggregation numbers and the observation of a slightly polydisperse relaxation mode in the autocorrelation scattering function may appear contradictory. This is not the case. It can be demonstrated that the relationship that links the aggregation number and the volume of a cluster transforms Eq. 1 into a distribution in size which exhibits a clear maximum. The diameter at the maximum is proportional to $\overline{N}_n$. It is thus important to differentiate here pfd's in aggregation number and in sizes as being distinct.

The observation of a rather broad polydispersity for the cluster size deserves some comments. This could be due in a first place to the intrinsic polydispersity of the cerium or of the iron oxide nanoparticles. The present results are for instance different from those obtained with surfactant micelles (complexed in the same conditions) for which the aggregation numbers were found to be narrowly distributed, with a polydispersity of 1.2 [11]. Experimental work is underway in order to verify the role of the particle polydispersity.

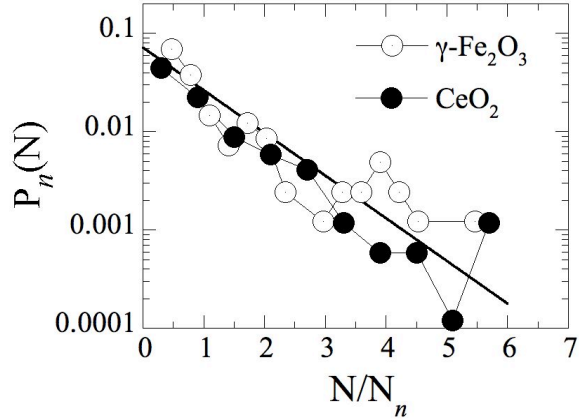

*Figure 4* : *Probability distribution function of reduced aggregation numbers $N/\overline{N}_n$ obtained for iron oxide and cerium oxide nanoparticle clusters. The statistics was established from a set of 200 aggregates for each system. The number of particles per cluster are $\overline{N}_n = 25 \pm 4$ (CeO$_2$) and $16 \pm 3$ ($\gamma$-Fe$_2$O$_3$). The straight line is calculated according to Eq. 1.*

Another assumption anticipates that the specific shape of the pdf's (Eq. 1) is an indication for the mechanism of formation of the mixed polymer-nanoparticle aggregates. It is striking to note that there is a straightforward analogy between the present results and those found in surfactant solutions of wormlike micelles. In wormlike micelles, the self-assembly is spontaneous and the aggregation number distribution is an exponentially decreasing function [24,25]. The linear growth of the aggregates is explained in terms of an excess curvature energy of the amphiphiles located in the end-caps. Using geometrical arguments, surfactant are predicted to spontaneously associate into elongated micelles when the critical packing parameter $V/a_0 \ell_C$ is comprised between 1/3 and 1/2 [26]. In the previous expression, V is the volume of a molecule, $\ell_C$ its length and $a_0$ the area of the polar head. Drawing the analogy further, we may now assume that the elementary building blocks in the complex formation can be described similarly, *i.e.* in terms of a critical packing ratio that takes into account the associating and interfacial constituents. In this model, a building block would comprise one nanoparticle and several copolymers, of the order of 10 in the present cases. A way to verify this assumption could be to modify the associating conditions, such as the molecular weight of the copolymers or the concentration.

# IV - Conclusion

In conclusion, we have found that for two systems of nanoparticles, the electrostatic complexation using polyelectrolyte-neutral copolymers give very similar results, namely the formation of stable nanoparticle clusters. An important result of the investigation is the direct visualization of clusters by electron microscopy and the evidence of the dense packing in the cores. This work suggests that controlled complexation with copolymers could be exploited in aqueous media to overcome the intrinsic instability of inorganic nanoparticles and for functionalization purposes.

**Acknowledgements** : We thank Bernard Cabane and Serge Stoll for fruitful discussions. The Laboratoire d'Electrochimie et de Chimie Analytique (UMR 7575 CNRS-ENSCP-UPMC) is kindly acknowledged for allowing us to use their Nanosizer Nano ZS for the solution characterization. This research is supported by Rhodia and by the Centre National de la Recherche Scientifique.

# References

[1] K. Kataoka; H. Togawa; A. Harada; K. Yasugi; T. Matsumoto; S. Katayose. Macromolecules 29 (1996) 8556 .
[2] A. Harada; K. Kataoka. Science 283 (1999) 65 .
[3] A. Harada; K. Kataoka. Macromolecules 31 (1998) 288 .
[4] J.-F. Berret; G. Cristobal; P. Hervé; J. Oberdisse; I. Grillo. Eur. Phys. J. E 9 (2002) 301 .
[5] S.v.d. Burgh; A.d. Keizer; M.A. Cohen-Stuart. Langmuir 20 (2004) 1073 .
[6] J.H. Jeong; S.W. Kim; T.G. Park. Bioconjugate Chem. 14 (2003) 473 .
[7] E. Raspaud; M. Olvera-de-la-Cruz; J.-L. Sikorav; F. Livolant. Biophys. J. 74 (1998) 381 .
[8] F. Bouyer; C. Gérardin; F. Fajula; J.-L. Puteaux; T. Chopin. Colloids Surf. A 217 (2003) 179 .
[9] T.K. Bronich; A.V. Kabanov; V.A. Kabanov; K. Yui; A. Eisenberg. Macromolecules 30 (1997) 3519 .
[10] J.-F. Berret; P. Hervé; O. Aguerre-Chariol; J. Oberdisse. J. Phys. Chem. B 107 (2003) 8111 .
[11] J.-F. Berret. J. Chem. Phys. 123 (2005) 164703.
[12] J.N. Cha; H. Birkedal; L.E. Euliss; M.H. Bartl; M.S. Wong; T.J. Deming; G.D. Stucky. J. Am. Chem. Soc. 125 (2003) 8285 .
[13] L.E. Euliss; S.G. Grancharov; S. O'Brien; T.J. Deminq; G.D. Stucky; C.B. Murray; G.A. Held. Nanoletters 3 (2003) 1489 .
[14] A. Ditsch; P.E. Laibinis; D.I.C. Wang; T.A. Hatton. Langmuir 21 (2005) 6006 .
[15] A. Sehgal; Y. Lalatonne; J.-F. Berret; M. Morvan. Langmuir 21 (2005) 9359 .
[16] J.K.G. Dhont. An Introduction to Dynamics of Colloids; Elsevier: Amsterdam, 1996.
[17] J.-P. Jolivet; R. Massart; J.-M. Fruchart. Nouv. J. Chim. 7 (1983) 325 .
[18] R. Massart; E. Dubois; V. Cabuil; E. Hasmonay. J. Magn. Magn. Mat. 149 (1995) 1 .
[19] J.-C. Bacri; R. Perzynski; D. Salin. J. Magn. Magn. Mat. 62 (1986) 36 .
[20] J.-F. Berret; N. Schonbeck; F. Gazeau; D.E. Kharrat; O. Sandre; A. Vacher; M. Airiau. J. Am. Chem. Soc. 128 (2006) 1555 .
[21] M. Destarac; W. Bzducha; D. Taton; I. Gauthier-Gillaizeau; S.Z. Zard. Macromol. Rapid Commun. 23 (2002) 1049 .
[22] J.-F. Berret; B. Vigolo; R. Eng; P. Hervé; I. Grillo; L. Yang. Macromolecules 37 (2004) 4922 .
[23] J.-F. Berret; K. Yokota; M. Morvan. Soft Materials 2 (2004) 71 – 84.
[24] M.E. Cates; S.J. Candau. J. Phys.: Condens. Matter 2 (1990) 6869 .
[25] J.-F. Berret. In Molecular Gels : Materials with Self-Assembled Fibrillar Networks; R.G. Weiss, P. T., Ed.; Springer, (2006).
[26] J.N. Israelachvili. Intermolecular and Surfaces Forces; Academic Press: New York, 1992.